\documentclass[prb,notitlepage,superscriptaddress]{revtex4-2}
\usepackage[T1]{fontenc}
\usepackage{textcomp}
\usepackage{amsmath}
\usepackage{amssymb}
\usepackage{graphicx}
\usepackage{esint}

\begin{document}

\title{Spatial Characterization of Fraunhofer Diffraction in a Four-Level Light-Matter Coupling System}% Force line breaks with \\

\author{Seyyed Hossein Asadpour}
 \affiliation {School of Physics, Institute for Research in Fundamental Sciences, IPM, Tehran, 19395-5531, Iran
 }%
 \author{Teodora Kirova}
 \email{teo@lu.lv}
\affiliation{Institute of Atomic Physics and Spectroscopy, University of Latvia, LV-1004, Latvia
}%
\author{Hamid R. Hamedi}
\affiliation{Institute of Theoretical Physics and Astronomy, Vilnius University, LT-10257, Lithuania
}%
\author{Reza Asgari}
\email{asgari@ipm.ir}
 \affiliation {School of Physics, Institute for Research in Fundamental Sciences, IPM, Tehran, 19395-5531, Iran
 }%
 
\date{\today}% It is always \today, today,

\begin{abstract}
We explore the spatial features of various orders of Fraunhofer diffraction patterns in a four-level $N$-type atomic system. The system interacts with a weak probe light, a standing wave (SW) coupling field in the $x$-direction, and a cylindrical beam of composite optical vortex type. We derive the first-order linear and third-order cross-Kerr nonlinear parts of the probe susceptibility by expanding the probe susceptibility of the system into the second order of the SW beam. This allows us to solve the integral equation of Fraunhofer diffraction, decoding its varying degrees to specific degrees of Bessel functions containing the nonlinear susceptibility. Notably, the nonlinear susceptibility exhibits dependence on the Orbital Angular Momentum (OAM) of the light beam, leading to spatial variations in the Bessel functions and, consequently, in the different orders of Fraunhofer diffraction. Leveraging the manipulation of OAM, we achieve precise control over the spatial mapping of diverse diffraction orders at various locations. Our research sheds new light on the spatial behavior of Fraunhofer diffraction in complex atomic systems. It presents exciting prospects for harnessing the OAM characteristics of light in future optical technologies.
\end{abstract}

%\keywords{Suggested keywords}%Use show keys class option if keyword
\maketitle

%\tableofcontents

\section{\label{sec:intro}Introduction}
Electromagnetically Induced Grating (EIG) is a fascinating phenomenon that occurs in atomic systems under the influence of laser fields. The concept of Electromagnetically Induced Grating (EIG) \cite{Ling1998, Mitsunaga1999, Cardoso2002} comes to the forefront when the conventional traveling wave (TW) coupling field, commonly observed in Electromagnetically Induced Transparency (EIT) systems \cite{Harris1997, Fleischhauer2005, Harris1990, Lukin2000}, is substituted with a standing-wave (SW) coupling beam. Upon introducing the SW control field, an intriguing spatial periodicity emerges in the absorption and dispersion characteristics of the TW probe beam, leading to the deflection of the probe field into high-order directions.
These unique EIG properties have significant implications for various applications in the field of optics and open up new perspectives on the interplay between light and matter and pave the way for the creation of optical systems and devices of the utmost quality. They find use in optical switching \cite{Brown2005}, storage of light \cite{Moretti2010}, all-optical beam splitting and fanning \cite{Zhao2010}, as well as novel implementations in electromagnetically induced Talbot effect \cite{Wen2017}, optical bistabilities \cite{Zhai2001}, topological insulators \cite{Zhang2015}, and soliton physics \cite{Zhang2011, Zhang2013}, to name a few.  
%_____________________________
\begin{figure}[b]
\includegraphics[width=1.0\textwidth]{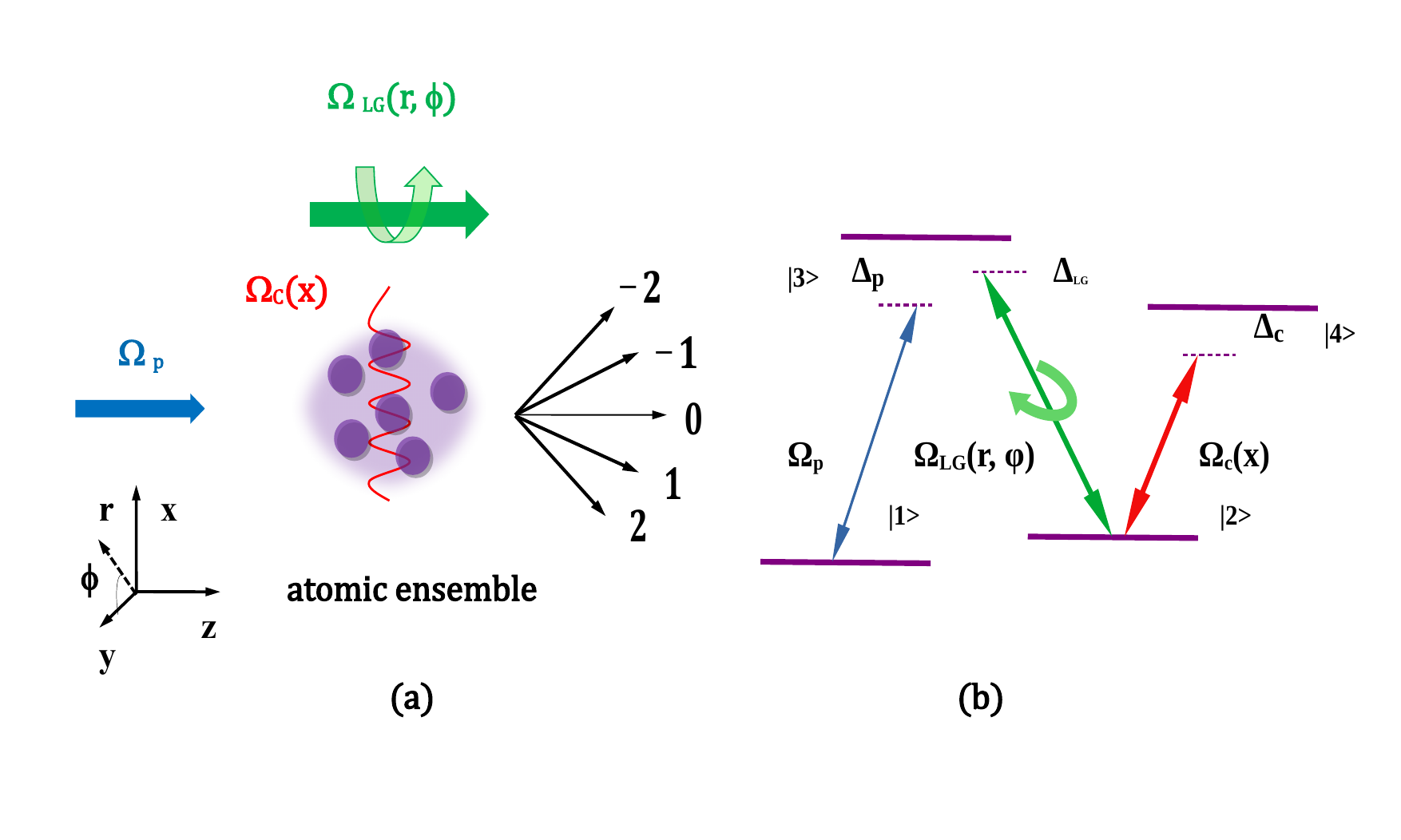}
\caption{\label{fig:scheme} (a) Excitation scheme of atomic ensemble in EIG; (b)} Proposed four-level $N$-type excitation scheme, with two ground states denoted as $|1\rangle$ and $|2\rangle$, along with two excited states represented by $|3\rangle$ and $|4\rangle$. These energy levels interact with three lasers: a weak TW probe field with Rabi frequency $\Omega_p$, a strong SW field with Rabi frequency $\Omega_{c}(x)$, and a Laguerre-Gaussian beam with Rabi frequency denoted as $\Omega_{LF}(r,\varphi)$.
\end{figure}
%________________________________
\\
\indent Current studies on the EIG have surpassed the traditional $\Lambda$-configuration and expanded to encompass interactions of many-level schemes with multiple beams, such as $\Xi$-type \cite{Dutta2006} and $Y$-type excitations \cite{Naseri2014}, tripod systems \cite{Wang2014}, and other four-level arrangements \cite{Sahrai2016}.
Furthermore, EIG effects have been successfully extended to various media, moving beyond the initial observations in hot and ultracold atoms. Notably, researchers have explored EIG in quantum dot molecules \cite{Hu2020, Feili2022, Liu2022}, showcasing its applicability in more complex systems. Additionally, we have recently described EIG in the vicinity of a plasmonic nanostructure in our works \cite{Asadpour2018, Asadpour2023}, highlighting its potential in novel applications involving plasmonics.
\\
\indent As emphasized in \cite{Asadpour2021}, the interaction of the EIG systems with optical vortex beams, i.e. light beams carrying Orbital Angular Momentum (OAM) \cite{Allen1999, Padgett2004, Babiker2019}, is especially beneficial in terms of providing new degrees of freedom for manipulating the performance of the optical grating.
In particular, it has been shown that by varying the winding number of the vortex beam, one can achieve direct control of the transfer of the probe beam energy from the zero- to high-diffraction orders \cite{Asadpour2021}.
\\
\indent The interaction of Laguerre-Gaussian (LG) beams with atomic systems represents a well-established and thriving direction in optical physics, leading to an increasing number of significant applications \cite{Dutton2004, Hamedi2019A, Ruseckas2011, Wang2008}. These applications span across diverse fields, including interdisciplinary studies \cite{Pugatch2007, Moretti2009}, further amplifying the impact of this research area.

Remarkably, recent advancements in experimental techniques have enabled the realization of optical vortex fields in various forms and complexities. Researchers can now engineer these fields with specific OAM, polarization, and in different geometrical arrangements.
\\
\indent Of particular interest to our research is the interaction of EIGs with composite optical vortices, which result from the interference of two or more vortex beams \cite{Maleev2003}. Under certain conditions for the OAMs, intriguing "petals"-like periodic intensity structures can be observed \cite{Baumann2009, Hamedi2019B}.
The ability to control the diffraction intensity distribution of the probe field to different locations in a two-dimensional (2D) EIG by manipulating the OAM of the composite vortex beam has been demonstrated in \cite{Asadpour2022}. This showcases the potential of tailored optical vortices in directing and shaping EIGs for desired applications.
Similarly, analogous effects were observed in a 2D EIG system utilizing a tripod configuration of atomic levels \cite{Wahab2023}. 
These results highlight the adaptability and importance of investigating composite optical vortices in EIG studies, where the exact regulation of light-matter interactions holds promise for a variety of prospective applications. 
\\
\indent EIGs also play a significant role in the generation of holographic gratings and optical modes. In a recent study by Arkhipkin et al. \cite{Arkhipkin2023}, the Fresnel diffraction of optical vortex light within a Raman interaction atomic medium with a spatially periodic pump field was investigated. Under specific conditions and at distances corresponding to classical Talbot planes, the study revealed the emergence of periodic amplitude-phase distributions. Another investigation \cite{Ikonnikov2023} focused on the near-field diffraction of optical vortex light on a fork-shaped grating with different topological charges. This study explored the evolution of specific optical vortices during propagation. 
\\
\indent Among other novel applications of composite Laguerre-Gaussian beams are bright/dark ring lattices for trapping atoms \cite{Franke2007, He2009}, including the creation of artificial gauge magnetic and electric fields \cite{Lembessis2017}, as well as more exotic phenomena, such as applications as wave-front sensors \cite{Blanchard2000}, or the so-called azimuths \cite{Desyatnikov2005, Bekshaev2006},  which represent a special class of spatially localized self-trapped ringlike beams.
\\
\indent In this current study, we expand upon our previous research \cite{Asadpour2022}, focusing on investigating the spatially dependent diffraction efficiencies of an $N$-type four-level EIG scheme interacting with composite vortex light. 
For this purpose, we employ two spatially-dependent coupling fields, a SW in the $x$-direction, and a composite vortex beam, both simultaneously interacting with the atomic energy levels. 
Expanding the steady-state probe susceptibility to the second order with respect to the SW beam allows us to uncover the contributions of its first-order linear effect and third-order cross-Kerr nonlinear effect. To analyze the resulting diffraction pattern, by employing analytical methods, we solve the integral equation of Fraunhofer diffraction. 
This solution reveals a dependence on Bessel functions that incorporate the nonlinear susceptibility. As the nonlinear susceptibility is contingent on the topological charge of the composite beam, it introduces spatial variations in the Bessel functions, and hence, subsequently, on different orders of Fraunhofer diffraction.
\indent In our numerical calculations, we first consider the cases of a simple vortex field, as well as a composite vortex with different winding numbers, subsequently we focus on the situation when a composite vortex beam with the same OAM numbers is applied with two specific cases for the detunings of the three laser fields interacting with the atomic transitions. 
Through these calculations, we demonstrate that the intensity pattern of the probe field diffraction exhibits spatial distribution at different locations, which can be effectively controlled by simply adjusting the OAM values of the composite vortex field. 
For zero values of the winding number, the diffraction of the zeroth, first, and second orders display distinct ring patterns with varying intensities, arising from destructive and constructive interference. 
When the OAM is nonzero, these ring shapes transform into intriguing petal-like structures, diminishing in size as the diffraction order grows. 
In addition, we study the behavior of the different orders of Fraunhofer diffraction patterns versus the atomic interaction length for different OAM numbers, showing how the energy of the probe field is distributed.
\\
\indent The simplicity and effectiveness of this proposed scheme for controlling diffraction intensity make it particularly appealing for future experimental realizations and applications. 
It offers a mechanism to customise and manipulate light-matter interactions to suit particular applications in optical technologies, which has significant potential for creating photonic components and systems that take advantage of the OAM of light.
\\
\indent The paper is organized along the following lines. 
After an Introduction, in Section~\ref{sec:model} we develop the groundwork of our theoretical model for the planned numerical calculations.
Employing Maxwell’s equation in the slowly varying envelope approximation in the steady state regime, as well as the Fraunhofer diffraction equation, in Section~\ref{sec:Fraun} we proceed to derive the analytical expressions for the spatial diffraction intensities of the zero-, first-, and second orders.
Section~\ref{sec:results} is devoted to the analysis of the obtained numerical results of the spatial diffraction intensities, including a variation of different system parameters, such as laser beam detunings and field strengths, atomic interaction length, as well as OAM numbers same or different values of the composite vortex beam.
In Section~\ref{sec:conclusions} we recap the main results, in parallel giving directions for future experimental realizations and feasible applications of our studies.
%_________________________________________________________
%_________________________________________________________
%_________________________________________________________
\section{\label{sec:model}Model and Theory}
\indent The system of interest consists of in a four-level $N$-type atomic configuration with two ground states $|1\rangle$, $|2\rangle$ and two excited states $|3\rangle$, $|4\rangle$  as shown in Fig.~\ref{fig:scheme}.
The atomic levels in our system are coupled by three laser fields. The first laser is a weak light with a Rabi frequency of $\Omega_p$, which specifically probes the $|1\rangle \rightarrow |3\rangle$ transition. The transition the $|2\rangle \rightarrow |4\rangle$ is derived with a coupling field with Rabi frequency of $\Omega_c$. Finally, a third laser beam, denoted as $\Omega_{LG}$ interacts with the $|2\rangle \rightarrow |3\rangle$ transition.  
The Rabi-freqiencies of probe and coupling fields are defined as $\Omega_{p}=\vec{\mu}_{13}\cdot \vec{E}_{p}/2\hbar$, $\Omega_{c}=\vec{\mu}_{24}\cdot \vec{E}_{c}/2\hbar$ and $\Omega_{LG}=\vec{\mu}_{23}\cdot \vec{E}_{LG}/2\hbar$,  with $\vec{\mu}_{ij}$ being the electric-dipole transition matrix element of $|i\rangle \rightarrow |j\rangle$.
\\
\indent  We adopt the interaction picture, employing the electric dipole and rotating wave Approximations. Under these assumptions, the total Hamiltonian of the system can be expressed as follows:
\begin{eqnarray}
H_{int}=&&-\hbar (\Delta_{p}-\Delta_{LG} )| 2 \rangle \langle 2 |-\hbar \Delta_{p} |3\rangle \langle 3 |
-\hbar (\Delta_{p}-\Delta_{LG} + \Delta_{c}) |4 \rangle \langle 4|\nonumber\\
&& -\hbar(\Omega_{p}|3\rangle \langle 1| +\Omega_{LG}|3\rangle \langle 2|+\Omega_{c}|4\rangle \langle 2| +H.C.).
\label{eq:Ham}
\end{eqnarray}
\indent In the above, we define the laser detunings as $\Delta_{p}=\omega_{p}-\omega_{31}$, $\Delta_{c}=\omega_{c}-\omega_{42}$, and $\Delta_{LG}=\omega_{LG}-\omega_{32}$, where $\omega_{p}$, $\omega_{c}$, $\omega_{LG}$ are the laser beams frequencies, and $\omega_{ij}, (i, j=1..4)$ are the resonant frequencies of the corresponding atomic transitions $|i\rangle \rightarrow |j\rangle$.
\\
\indent  The wave function of the system is decomposed into the basis set of atomic levels;  $\{|1\rangle,|2\rangle,|3\rangle,|4\rangle\}$, i.e.  $\Psi(t)=\sum_{i=1,4}{a_{i}(t)e^{-i\omega_{i}t}|i\rangle}$, involving the time-dependent coefficients $a_{i}(t)$, as well as the atomic levels energies $\hbar\omega_{i}, (i=1...4)$.
The dynamics of the system are described by the equations of motion for the probability amplitudes of all atomic states:
\begin{eqnarray}
\frac{d {a_{1}}}{dt} &=& i\Omega_{p} a_{3},\nonumber\\
\frac{d {a_{2}}}{dt} &=& i\left[ (\Delta_{p}-\Delta_{LG})a_{2}+ \Omega_{LG} a_{3}+\Omega_{c} a_{4}\right]-\gamma_{2} a_{2},\nonumber\\
\frac{d {a_{3}}}{dt} &=& i(\Omega_{p} a_{1}+\Omega_{LG} a_{2}+\Delta_{p} a_{3})-\frac{\Gamma_{3}}{2} a_{3},\nonumber\\
\frac{d {a_{4}}}{dt} &=& i\left[ \Omega_{c} a_{2}+(\Delta_{p}-\Delta_{LG} +\Delta_{c})a_{4}\right]-\frac{\Gamma_{4}}{2}a_{4}.
\label{eq:Amp}
\end{eqnarray}
\indent 
The parameters $\Gamma_3$, $\Gamma_4$, and $\gamma_2$ indicate the decay rates from the upper levels $|3\rangle$, and $|4\rangle$, as well as the ground state decoherence, respectively. 
In what follows, we will assume $\Gamma_{3}=\Gamma_{4}=\gamma$ and express all other parameters in units of $\gamma$.
\\
\indent Under the weak probe field approximation, specifically when $|a_{1}|^{2}\approx1$, the susceptibility of the probe field can be derived by solving Eqs.~(\ref{eq:Amp}) under the steady-state condition as follows:
\begin{equation}
\chi_{p}=\frac{N\mu_{13}^2}{2\varepsilon_{0}\hbar}\chi(\omega_{p}),
\label{eq:ChiP}
\end{equation}
\noindent 
We use the definition of the polarization of the medium $\vec{P}_{p}=N\vec{\mu}_{13}a_{3}a_{1}^{*}$ where  $N$ shows the atomic density, and $\varepsilon_{0}$ is the dielectric constant in vacuum.
The form of $\chi(\omega_{p})$ is given by
\begin{equation}
\chi(\omega_{p})=\frac{A_{2}A_{4}-\Omega_{c}^{2}}{A_{4}(A_{2}A_{3}-\Omega_{LG}^{2})-A_{3}\Omega_{c}^{2}},
\label{eq:Chi}
\end{equation}
\noindent where, for the seek of simplicity, we  introduce the following notations:
\begin{eqnarray}
A_{2}=\Delta_{p} -\Delta_{LG}+i\gamma_{2};
A_{3}=\Delta_{p}+i\frac{\Gamma_{3}}{2}; 
A_{4}=\Delta_{p}-\Delta_{LG}+\Delta_{c}+i\frac{\Gamma_{4}}{2}.
\label{eq:As}
\end{eqnarray}
\\
\indent To analyze the nonlinear modulation induced by the control field with SW pattern, we expand $\chi(\omega_{p})$ into the second order of $\Omega_{c}$ as
\begin{equation}
\chi(\omega_{p})=\chi^{(1)}(\omega_{p})+\Omega_{c}^{2}\chi^{(3)}(\omega_{p}).
\label{eq:Chi13}
\end{equation}
The first and third-order Kerr nonlinear parts of the probe susceptibility are then given by
\begin{eqnarray}
\chi^{(1)}(\omega_{p})=-\frac{A_{3}}{A_{2}A_{3}-\Omega_{LG}^{2}(r,\varphi)},\\
\chi^{(3)}(\omega_{p})=-\frac{\Omega_{LG}^{2}(r,\varphi)}{A_{4}|A_{2}A_{3}-\Omega_{LG}^{2}(r,\varphi)|^2}.
\label{eq:Chi1Chi3}
\end{eqnarray}
\indent Control fields that have SW patterns in turn cause spatial modulation of the probe beam absorption and refraction. 
As a consequence of the intensity-dependent susceptibility, the atomic system acts as a grating, diffracting the probe beam in different directions.
\\
\indent For the purposes of observing these effects, we will then utilize a coupling field $\Omega_{c}=\Omega_{c0} [ \sin(\pi x/\Lambda_{x})]$, which constitutes a SW with a space frequency $\Lambda_{x}$ along the $x$ direction.
Moreover, we replace the $\Omega_{LG}$ field with a composite vortex beam, i.e. a superposition of two vortices: 
\begin{equation}
\Omega_{LG}=\Omega e^{-r^{2}/w^{2}} \left [(r/w)^{ \lvert l_{1} \rvert} e^{il_{1}\varphi} +(r/w)^{ \lvert l_{2} \rvert} e^{il_{2}\varphi}\right ],
\label{eq:OmegaLG}
\end{equation}
where $w$ is the beam waist and its value is in order of $\mu m$, the radial distance from the axis of the LG beam is represented by $r=\sqrt{x_1^2+y_1^2}$, $l_{1}, l_{2}$ provide the winding numbers, and $\varphi$ is the azimuthal angle.
We can simplify the expressions by assuming $l_{1}=-l_{2}=l$:
\begin{equation}
\Omega_{LG}=\Omega_{lg}\cos(l\varphi);
\Omega_{lg}=2\Omega e^{-r^{2}/w^{2}}(r/w)^{\lvert l \rvert}.
\label{eq:OmegaLG1}
\end{equation}
%_________________________________________________________
%_________________________________________________________
%_________________________________________________________
\section{\label{sec:Fraun}Fraunhofer Diffraction Pattern}
By employing Maxwell's equations in the slowly varying envelope approximation and considering the steady-state regime, we can derive the diffraction pattern of the probe beam as follows:
\begin{eqnarray}
\frac{\partial E_{p}}{\partial z}=i\frac{\pi}{\varepsilon_{0}\lambda_{p}}P_{p}, \quad
P_{p}=\varepsilon_{0}\chi (\omega_{p})E_{p},
\label{eq:Maxw}
\end{eqnarray}
\noindent with $\lambda_{p}$ denoting the probe light wavelength. 
The above can be rewritten as:
\begin{equation}
\frac{\partial E_{p}}{\partial z^{'}}=i\gamma\chi(\omega_{p})E_{p},
\end{equation}
\noindent where we have introduced the definition $z'=(\pi N \mu_{13}^{2})/(\hbar \epsilon_{0}\lambda_{p}\gamma)z$.
In what follows we will work with a dimensionless $z'$, by treating $\xi=(\hbar \epsilon_{0}\lambda_{p}\gamma)/(\pi N \mu_{13}^{2})$ as the unit for $z$. The parameter $\xi$ can be controlled by the atomic density of the medium, and its order is about $\mu m$.
\\
\indent The transmission function of the grating is determined by: 
\begin{equation}
T(x)=e^{-Im(\chi(\omega_{p}))L}e^{iRe(\chi(\omega_{p}))L},
\label{eq:Tx}
\end{equation}
\noindent where $L$ is the interaction length. The first and second exponential terms correspond to the grating amplitude and phase modulation, respectively.
After performing Fourier transforming of Eq. (\ref{eq:Tx}), we arrive at the expression for Fraunhofer diffraction:
\begin{widetext}
\begin{equation}
I_{p}(\theta_{x})=|E(\theta_{x})|^{2}\frac{\sin^{2}(M\pi \Lambda_{x}  \sin\theta_{x}/\lambda_{p})}{M^{2}\sin^{2}(\pi \Lambda_{x} \sin \theta_{x}/\lambda_{p})},
\label{eq:Ip}
\end{equation}
\end{widetext}
\noindent where 
\begin{equation}
E(\theta_{x})=\int_{0}^{1}T(x)\exp(-i2\pi x\Lambda_{x} \sin\theta_{x}/\lambda_{p})dx.
\label{eq:Etheta}
\end{equation}
The $m^{th}$ order diffraction angle $\theta_{x}$ with respect to the $z$ direction is calculated from $\sin \theta_{x}=m\lambda_{p}/\Lambda_{x}$.
\\
\indent We are interested in investigating the Fraunhofer diffraction orders by analytically solving Eq.~(\ref{eq: Etheta}).
For this purpose, we need to solve Eq.~(\ref{eq:Tx}) first.
By inserting Eq.~(\ref{eq:Chi13}) in Eq.~(\ref{eq:Tx}) the transmission function of the 2D grating is given by
\begin{equation}
T(x)=e^{-Im[\chi^{(1)}(\omega_{p})+\Omega_{c}^{2}\chi^{(3)}(\omega_{p})]L}e^{iRe[\chi^{(1)}(\omega_{p})+\Omega_{c}^{2}\chi^{(3)}(\omega_{p})]L}.
\label{eq:Tx1}
\end{equation}
It should be reminded that $
%\begin{equation}
\Omega_{c}^{2}(x)=\Omega_{c0}^{2} \sin^{2}(\pi x/\Lambda_{x})
\label{eq:Omegac2}
%\end{equation}
$. 
By presenting $\chi^{(3)}(\omega_{p})$ in the form of $
%\begin{equation}
\chi^{(3)}(\omega_{p})=A+iB; Re(\chi^{(3)}(\omega_{p}))=A; Im(\chi^{(3)}(\omega_{p}))=B
\label{eq:Chi3}
%\end{equation}
$, and using a new notation $k=\exp[(-Im(\chi^{(1)}(\omega_{p}))+iRe(\chi^{(1)}(\omega_{p})))L]$, we get a new expression of Eq.~(\ref{eq:Etheta}):
\begin{equation}
E(\theta_{x})= k\int_{0}^{1}\exp[iL\chi^{(3)}(\omega_{p}) \Omega_{c0}^{2} \sin^{2}(\pi x /\Lambda_{x})]\exp(-i2\pi x\Lambda_{x} \sin\theta_{x}/\lambda_{p})dx.
\label{eq:Etheta1}
\end{equation}
%Here we have used the notation:
%\begin{equation}
%k=\exp[(-Im(\chi^{(1)}(\omega_{p}))+iRe(\chi^{(1)}(\omega_{p})))L]
%\label{eq:k}
%\end{equation}
By setting $M=L \chi^{(3)}(\omega_{p})\Omega_{c0}^{2}$  in Eq.~(\ref{eq:Etheta1}) we obtain:
\begin{equation}
E(\theta_{x})= k\exp(\frac{iM}{2})\int_{0}^{1}\exp(-i\frac{M}{2}\cos(2\pi x/\Lambda_{x}))\exp(-i2\pi x\Lambda_{x} \sin\theta_{x}/\lambda_{p})dx.
\label{eq:Etheta2}
\end{equation}
%________________
%_______________
\begin{figure}[b]
\includegraphics[width=1\textwidth]{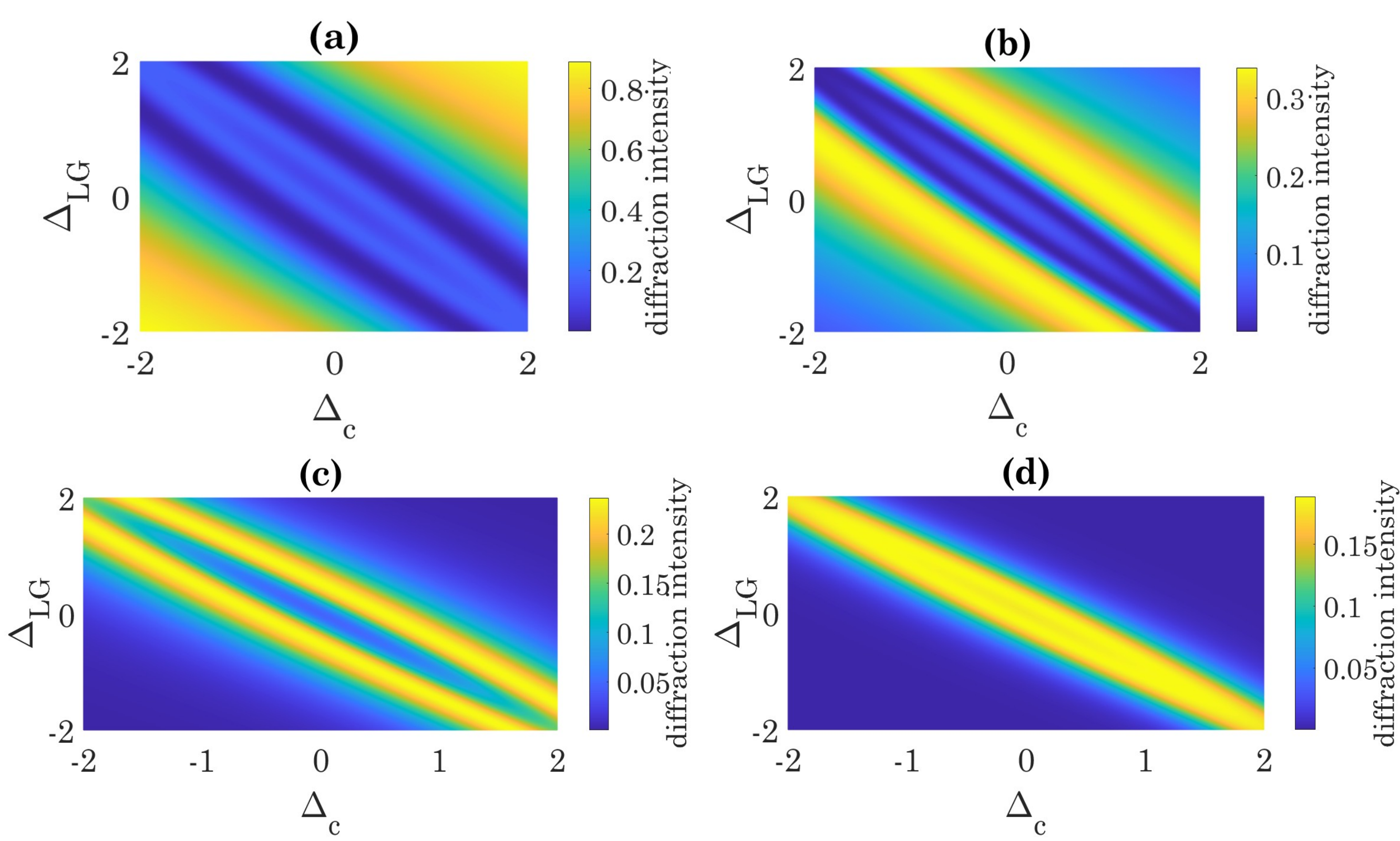}
\caption{\label{fig:Fig2} Different orders of the Fraunhofer diffraction pattern versus the parameters $\Delta_{c}$ and $\Delta_{LG}$. (a) corresponds to the zero-order, (b) corresponds to the first-order, (c) corresponds to the second-order, and (d) corresponds to the third-order. The selected parameters are $\Delta_{p}=0$, $\Omega_{c0}=0.5\gamma$, $\Omega=1.5\gamma$, $L/\xi=50$.}
\end{figure}
%________________
To further simplify the above expression we introduce $\phi=\frac{2\pi x}{\Lambda_{x}}$ and $n=\Lambda_{x}\sin\Theta_{x}/\lambda_{p}$ and arrive at:
\begin{equation}
E(\theta_{x})= k\exp(\frac{iM}{2})\int_{0}^{2\pi}\exp(-i\frac{M}{2}\cos\phi)\exp(-in\phi)\left(\frac{\Lambda_{x}}{2\pi}\right)d\phi.
\label{eq:Etheta3}
\end{equation}
Next, we define $M'=-M/2$ and $P=k\exp(\frac{iM}{2})\left(\frac{\Lambda_{x}}{2\pi}\right)$, and we use the following integral:
\begin{equation}
\int_{0}^{2\pi}\exp(-i(n\phi-M'\cos\phi))d\phi=
\int_{0}^{2\pi}(\cos(n\phi-M'\cos\phi)-i\sin(n\phi-M'\cos\phi))d\phi.
\label{eq:P}
\end{equation}
Since the first term of Eq.~(\ref{eq:P}) gives zero, Eq.~(\ref{eq:Etheta3}) attains to a  simplified form:
\begin{equation}
E(\theta_{x})=-iP\int_{0}^{2\pi}\sin(n\phi-M'\cos\phi)d\phi=iP2\pi J_{n}(M'),
\label{eq:Etheta4}
\end{equation}
\noindent where $J_{n}(M')$ represents a Bessel function of the $n^{th}$ order.
Plugging  the definitions of $P$, $M'$, and further $k$, $M$ into Eq.~(\ref{eq:Etheta4}), renders the final expression for $E(\theta_{x})$:
\begin{equation}
E(\theta_{x})= -i\Lambda_{x}\exp[iL(\chi^{(1)}(\omega_{p})+\frac{1}{2}\chi^{(3)}(\omega_{p})\Omega_{c0}^{2})]J_{n}\left(-\frac{L \chi^{(3)}(\omega_{p})\Omega_{c0}^{2}}{2}\right).
\label{eq:Ethetaf}
\end{equation}
We can now proceed to obtain the $n$-order diffraction intensity by combining Eq.~(\ref{eq:Etheta}) and Eq.~(\ref{eq:Ethetaf})
\begin{equation}
I(\theta_{x}^{(n)})=\bigg|\Pi J_{n}\left( -\frac{L \chi^{(3)}(\omega_{p})\Omega_{c0}^{2}}{2}\right)\bigg|^{2}.
\label{eq:Itheta}
\end{equation}
Here $\Pi$ stands for the following expression
\begin{equation}
\Pi=-i\Lambda_{x}\exp[iL(\chi^{(1)}(\omega_{p})+\frac{1}{2}\chi^{(3)}(\omega_{p})\Omega_{c0}^{2})]. 
\label{eq:Pi}
\end{equation}   
In the following discussion, we mainly concentrate on the spatial diffraction intensities of the zero-, first-, and second orders as follows:
\begin{eqnarray}
I(\theta_{x}^{(0)})=\bigg|\Pi J_{0}\left( -\frac{ L \chi^{(3)}(\omega_{p})\Omega_{c0}^{2}}{2}\right)\bigg|^{2},\nonumber \\
I(\theta_{x}^{(1)})=\bigg|\Pi J_{1}\left( -\frac{L \chi^{(3)}(\omega_{p})\Omega_{c0}^{2}}{2}\right)\bigg|^{2}, \nonumber \\
I(\theta_{x}^{(2)})=\bigg|\Pi J_{2}\left( -\frac{L \chi^{(3)}(\omega_{p})\Omega_{c0}^{2}}{2}\right)\bigg|^{2}.
\label{eq:Itheta}
\end{eqnarray}
It is clear from expressions in Eq.~(\ref{eq:Itheta}) that the different orders of the Fraunhofer diffraction are directly related to the identical orders of the Bessel function.
%_______________
\begin{figure}[b]
\includegraphics[width=1\textwidth]{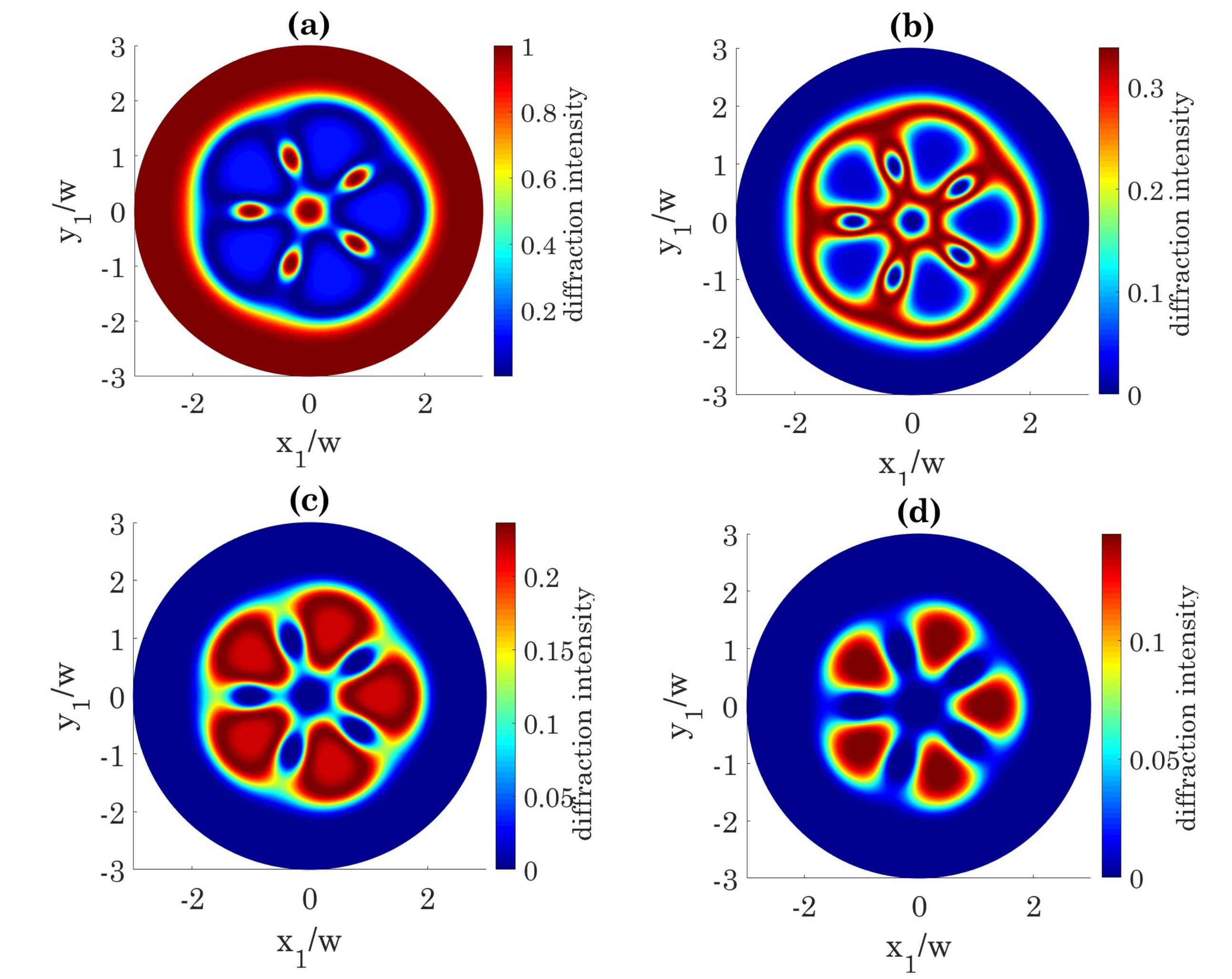}
\caption{\label{fig:Fig3} Spatial dependency of the different diffraction orders vs $x_{1}/w$ and $y_{1}/w$ for $l_{1}=4$ and $l_{2}=-1$. (a) corresponds to the zero-order, (b) corresponds to the first-order, (c) corresponds to the second-order, and (d) corresponds to the third-order. The selected parameters are
$\Delta_{p}=0$, $\Delta_{c}=-1$, $\Delta_{LG}=2$, $\Omega_{c0}=0.5\gamma$, $\Omega=1.1\gamma$, $L/\xi=50$.}
\end{figure}
%_______________
\begin{figure}[b]
\includegraphics[width=1\textwidth]{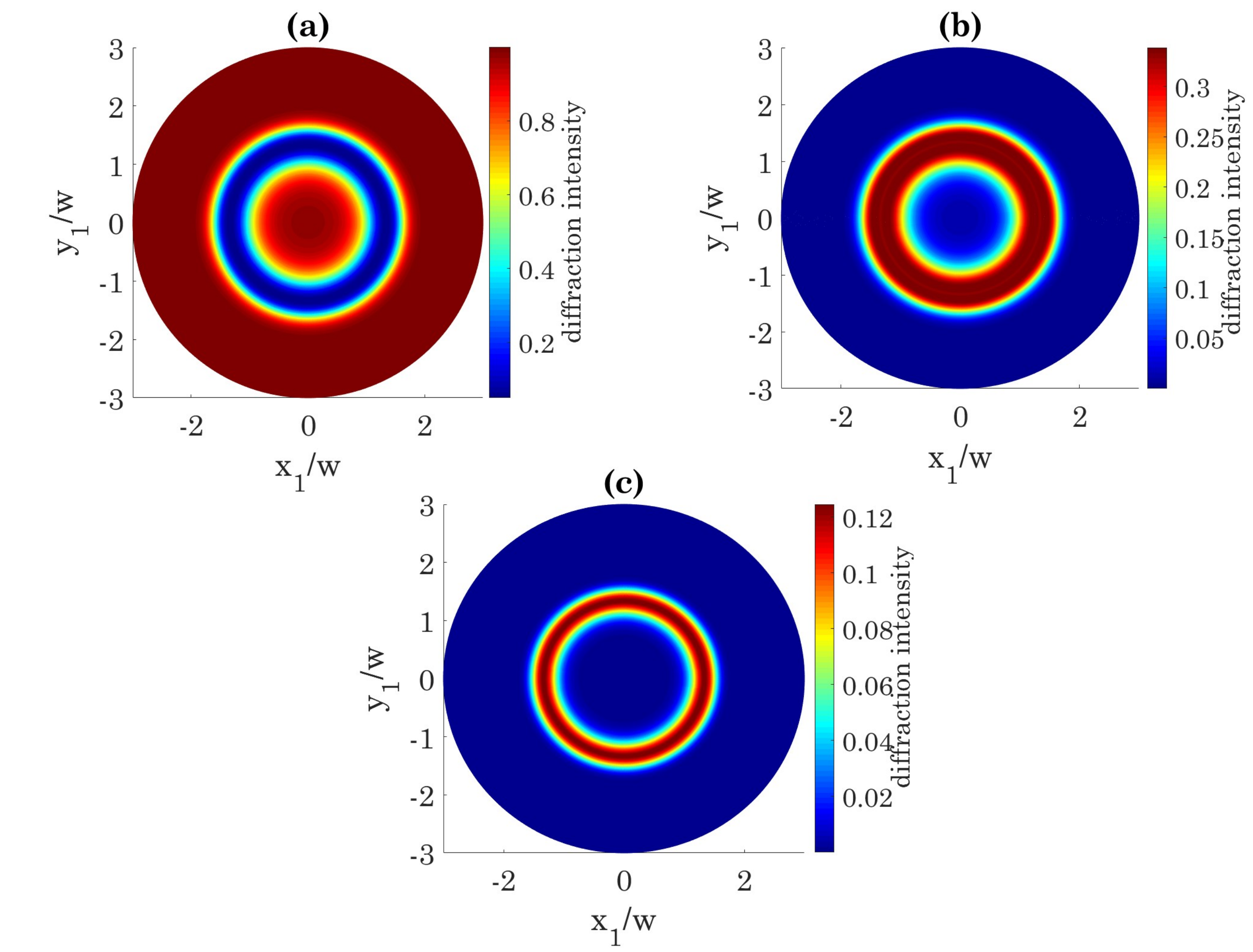}
\caption{\label{fig:Fig4}   Spatial dependency of the different diffraction orders vs $x_{1}/w$ and $y_{1}/w$ for winding number $l=0$. (a) corresponds to the zero-order, (b) corresponds to the first-order and (c) corresponds to the second-order. The selected parameters are
$\Delta_{p}=\Delta_{c}=\Delta_{LG}=0$, $\Omega_{c0}=0.2\gamma$, $\Omega=1.5\gamma$, $L/\xi=50$. Notice that for the first and second orders
of diffraction, the energy distribution undergoes a pattern in which the central region
experiences a complete absence of energy, resulting in a dark spot at the center.}
\end{figure}
%_______________
\begin{figure}[b]
\includegraphics[width=1\textwidth]{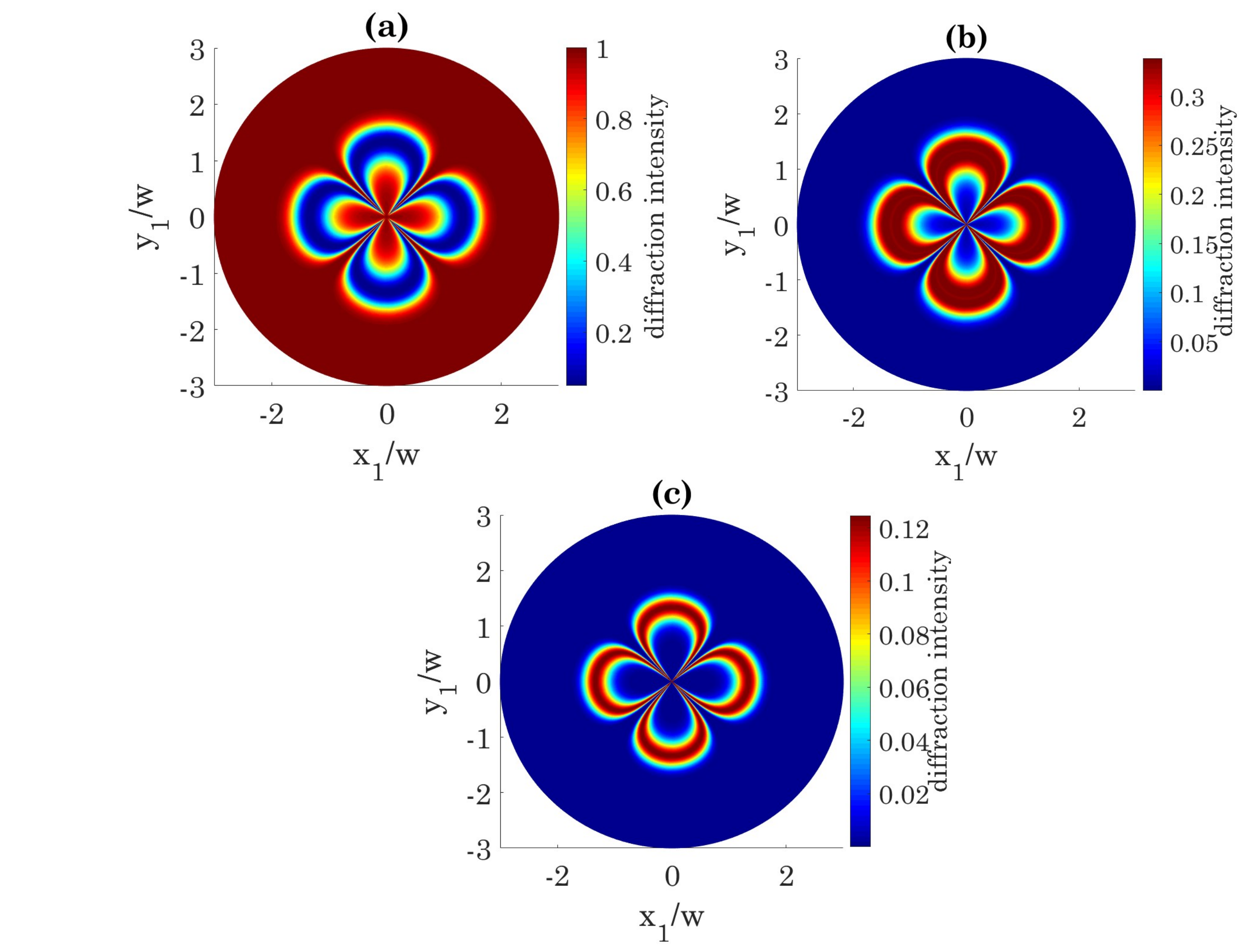}
\caption{\label{fig:Fig5}  Spatial dependency of the different diffraction orders vs $x_{1}/w$ and $y_{1}/w$ for winding number $l=2$. (a) corresponds to the zero-order, (b) corresponds to the first-order and (c) corresponds to the second-order. The light distribution within the azimuthal space differs, depending on the diffraction order. The selected parameters are the same as in Fig.~\ref{fig:Fig4}.}
\end{figure}
%______________________________________________________
%_______________________________________________________
%_______________________________________________________
\section{\label{sec:results}Results and Discussions}
\indent In this section, we will discuss the spatial dependency of the different orders of the grating by adjusting the OAM of the composite vortex light  $l$ in cases of different detunings of the three laser fields interacting with the $N$-type atomic system.
We will also investigate how the different orders of Fraunhofer diffraction behave, i.e. what is the distribution of probe field energy, as a function of the atomic interaction length for different winding numbers of the composite vortex beam.
%_____________________________________________
\subsection{\label{sec:simple}Case of simple vortex field}
\indent To begin with a simpler scenario, let us consider the setup where a basic vortex beam is employed. 
In Fig.~\ref{fig:Fig2}, we display the different orders of the Fraunhofer diffraction pattern versus the detunings of the coupling fields $\Delta_{c}$ and $\Delta_{LG}$ when the probe light is in resonance with its transition ( i.e. $\Delta_{p}=0$).
As seen here, the intensity of the different orders can be controlled by the detunings of the coupling lights.
When the coupling lights are off-resonate, more of the probe energy gathers in the first order ($-1 \le \Delta_{c} \le -2 $; $-1 \le \Delta_{LG} \le -2 $ ) and ($1 \le \Delta_{c} \le 2 $; $1 \le \Delta_{LG} \le 2 $). 
\indent In other similar regimes for the detunings (indicated by the yellow band in the higher orders) the intensity of the diffraction orders may be higher than the zero-order.
%_____________________________________________
%_____________________________________________
\subsection{\label{sec:composite_diff}Case of composite vortex with different OAMs}
\indent As a second step we apply the composite form of the optical vortex light (Eq.~(\ref{eq:OmegaLG})) considering different values of the winding numbers $l_{1}$ and $l_{2}$.
As shown in Fig.~\ref{fig:Fig3}, we present the Fraunhofer diffraction patterns for this composite beam with $l_{1}=4$ and $l_{2}=-1$.
The results in Fig.~\ref{fig:Fig3} are decomposed into different parts to visualize the various
diffraction orders. 
Parts (a), (b), (c), and (d) correspond to the zero order,  the first order, the second order, and the third order, respectively.
This comprehensive representation of the different diffraction orders provides a more complete and detailed view of the pattern for the composite beam.
\\
\indent As discussed in \cite{Hamedi2019A}, a composite twisted beam with different values of $l_{1}$ and $l_{2}$ exhibits a central vortex of charge $l_{1}$, surrounded by $|l_{1}-l_{2}|$ peripheral vortices.
Our findings concerning the diffraction patterns and the spatial mapping of different orders strongly align with the structural characteristics of such a composite vortex.
Specifically, we observe that the different diffraction orders of the grating exhibit a
prominent 5-fold symmetry, which directly corresponds to the presence of $|l_{1}-l_{2}|$
peripheral vortices in the composite beam. 
For the zero-order grating (Fig.~\ref{fig:Fig3}(a)), the majority of light is concentrated at the core of the azimuthal space, consistent with the central vortex of the composite beam. 
Furthermore, the zero-order grating exhibits five peripheral spots precisely located at the positions where the singularity points of the peripheral vortices in the composite beam are situated.
Surrounding this central region, there is a ring-shaped area where the probe field
energy is entirely reconstituted.
\indent In contrast, the first-order grating illustrated in Fig.~\ref{fig:Fig3}(b) produces a distinctive wheel-like pattern, indicating a significant change in energy distribution compared to the zero-order. 
In this case, the energy is concentrated in locations that were previously completely devoid of it, essentially creating an anti-phase relationship with the zero-order. 
With the higher second-(Fig.~\ref{fig:Fig3}(c)) and third-- (Fig.~\ref{fig:Fig3}(d)) orders of grating, the patterns undergo structural changes while still retaining the 5-fold symmetry. 
%_____________________________________________
%_____________________________________________
\subsection{\label{sec:composite_diff}Case of composite vortex with same OAMs}
\indent As mentioned in Section ~\ref{sec:model}, the main focus of our investigations is the case of equal winding numbers $l_{1}=l_{2}$ of the composite vortex coupling field ( see Eq. (\ref{eq:OmegaLG1})).
%__________________________________
\textbf{\subsubsection{\label{sec:detunings0}Resonant probe and coupling fields}}
We first investigate the case when all laser beams are resonant with the respective atomic transitions, e.g. $\Delta_{p}=\Delta_{c}=\Delta_{LG}=0$.
In Fig.~\ref{fig:Fig4}, we present the spatial properties of the different orders of diffraction patterns for a winding number $l=0$. 
The blue areas indicate regions of low intensity in the diffraction pattern, while the dark red structures represent positions of high-intensity diffraction.
When $l=0$, the diffraction intensities of the zero- (a), first- (b), and second- (c) orders exhibit distinct ring patterns. For the zero-order, the majority of the energy is concentrated at the center, forming a high-intensity region surrounded by a ring-shaped area where the probe field energy is completely absent. 
This ring-shaped void represents a region of destructive interference, arising from the diffraction process. Moving away from this zero-energy ring, the energy starts to reappear.
In contrast, for the first and second orders of diffraction, the energy distribution undergoes a fascinating transformation. In these cases, the central region experiences a complete absence of energy, resulting in a dark spot at the center. 
As one moves away from the center, the energy becomes manifest, taking the form of a ring,  indicating its presence. This ring-shaped region represents constructive interference, where the diffracted waves combine to create regions of increased intensity.
As the order of the grating increases, the ring patterns gradually shrink and narrow. 
This reduction in size and width of the grating rings is a consequence of the increasing complexity of the diffraction process, leading to more intricate interference patterns. 
In higher-order diffraction patterns, the ring-shaped regions become progressively smaller and tighter, reflecting the rich and intricate behavior of the diffraction process. Consequently, the number and intensity of these bright and dark regions depend on the size and microscopic structure of the atoms and the wavelength of the incident light.
\\
\indent Further, Fig.~\ref{fig:Fig5} shows the diffraction patterns and the spatial mapping of the various orders when $l=2$.
Here, we observe a petal-like pattern for the different diffraction orders. 
Figs.~\ref{fig:Fig5}(a)-(c) provide a visual representation of the diffracted patterns observed for the zero- to second-orders of diffraction. 
These figures reveal intriguing characteristics exhibited by gratings of different orders. 
Notably, each order of the grating displays a distinctive pattern with a symmetrical arrangement resembling petals. 
Upon closer examination, it becomes apparent that the light distribution within the azimuthal space differs, depending on the diffraction order. 
In the case of zero-order, the light is predominantly concentrated at the center, forming an intriguing anti-petal-like pattern. 
The energy of the diffracted light appears to converge towards the central region, resulting in reduced intensity towards the outer areas. 
However, as we move to higher orders of diffraction such as the first- and second-orders, the dynamics of light distribution undergoes a remarkable change. 
In these instances, the light is no longer focused at the center but gathers primarily at the petal regions. 
The central region, which was previously the focal point of energy, now experiences a complete absence of light. 
This unique behavior of light distribution creates a distinctive pattern with zero intensity at the center and enhanced intensity at the petals. 
Furthermore, as the order of diffraction increases, the petal patterns gradually diminish in size. 
This is clearly illustrated in Fig.~\ref{fig:Fig5}(c), where the petals become smaller and more compact, compared to the lower orders. 
The shrinking of the petal patterns with increasing order highlights the intricate relationship between diffraction and the resulting spatial distribution of light. 
%_____________________________________________
\textbf{\subsubsection{\label{sec:detunings1}Resonant probe and off-resonant coupling fields}}
\indent In what follows, we will study the azimuthal dependence of the different diffraction orders in the off-resonance conditions for the coupling and composite optical vortex lights, while the probe field is still resonant, i.e. $\Delta_{p}=0, \Delta_{c}=-\gamma, \Delta_{LG}=2\gamma$.
%_________________________
\begin{figure}[b]
\includegraphics[width=1\textwidth]{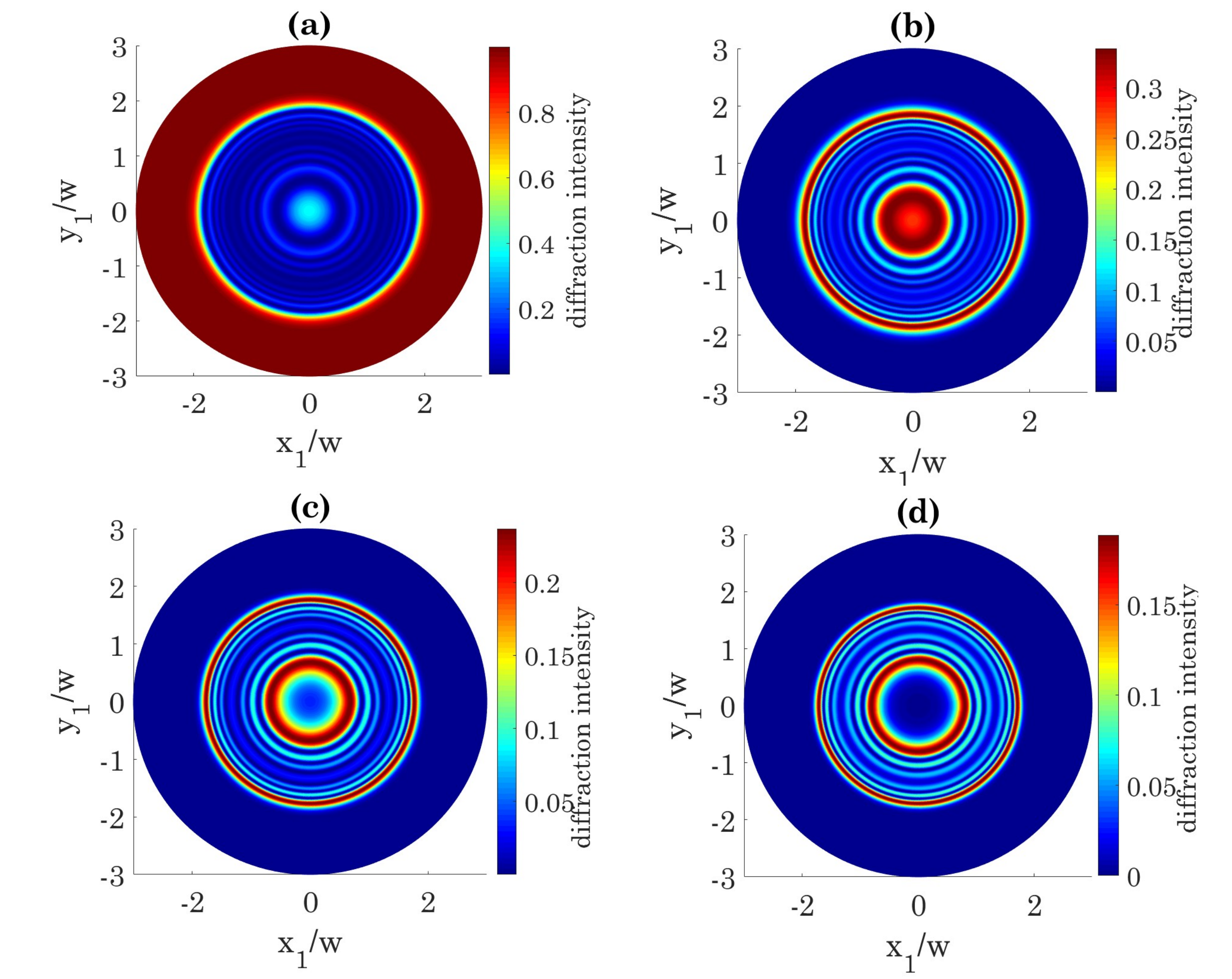}
\caption{\label{fig:Fig6}  Spatial dependency of the different diffraction orders vs $x_{1}/w$ and $y_{1}/w$ for winding number $l=0$. (a) corresponds to the zero-order, (b) corresponds to the first-order, (c) corresponds to the second-order, and (d) corresponds to the third-order. The selected parameters are $\Delta_{c}=-\gamma$, $\Delta_{LG}=2\gamma$, $\Omega_{c0}=0.5\gamma$, and the rest are the same as in Fig.~\ref{fig:Fig4}.}
\end{figure}
%____________________
\begin{figure}[b]
\includegraphics[width=1\textwidth]{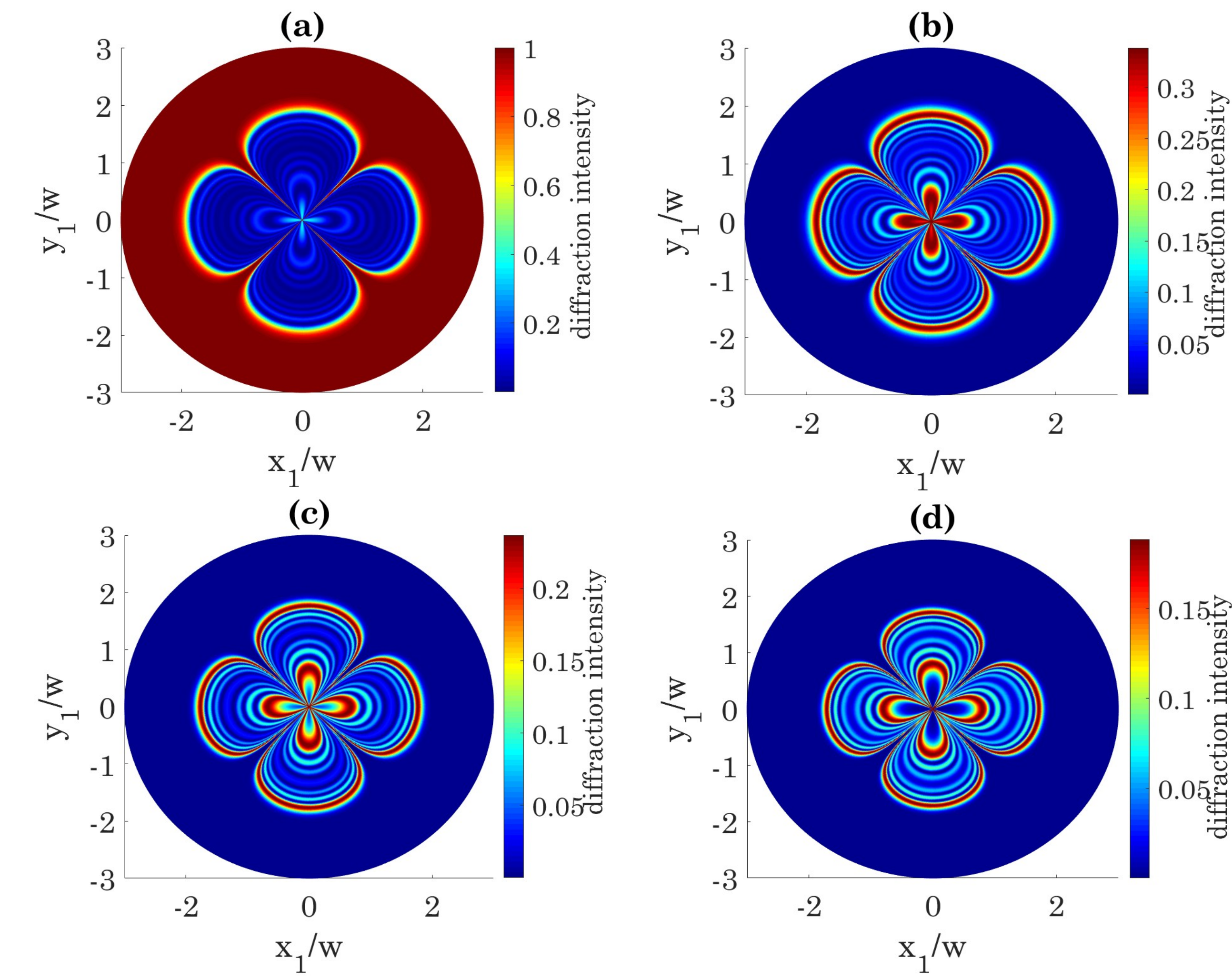}
\caption{\label{fig:Fig7}  Spatial dependency of the different diffraction orders vs $x_{1}/w$ and $y_{1}/w$ for winding number $l=2$. (a) corresponds to the zero-order, (b) corresponds to the first-order, (c) corresponds to the second-order, and (d) corresponds to the third-order. The intensity distribution exhibits petal-like patterns for different degrees of diffraction. The selected parameters are the same as in Fig.~\ref{fig:Fig6}.}
\end{figure}
%_______________________
\indent The numerical results for the case of winding number $l=0$ are given in Fig.~\ref{fig:Fig6}.
In this analysis, it is evident that the distribution of intensity in the zero-order (Fig.~\ref{fig:Fig6}(a)) attains its maximum value in regions that are situated away from the center. 
At the same time, the distribution of intensities experiences a lower value in the central regions. 
In the case of the first-order (Fig.~\ref{fig:Fig6}(b)), the intensity distribution reaches its maximum value at the center and gradually decreases in regions distant from the center.
The second-order (Fig.~\ref{fig:Fig6}(c)) reveals that the intensity distribution is visible in the central regions, but its value is comparably lower than that of the zero-and first-orders.
Nevertheless, in some narrow regions of space, the distribution of intensity surpasses the zero- and first-orders. 
As for the third-order, shown in Fig.~\ref{fig:Fig6}(d), we observe that there is no intensity distribution in both the central and outer regions, with only a few areas exhibiting a weak intensity distribution, which is inferior to that of other orders.
\\
\indent Furthermore, we explore the behavior of the grating for the nonzero OAM number $l=2$, as depicted in Fig.~\ref{fig:Fig7}. The intensity distribution exhibits petal-like patterns for different degrees of diffraction. Notably, the probe field's highest energy concentration occurs in the first-order (Fig.~\ref{fig:Fig7}(b)) within the central petal region. As one moves away from the center, the intensity distribution for other orders displays thin petal-shaped regions. 
%____________________________
\textbf{\subsubsection{\label{sec:Ldependence}Dependence on atomic interaction length}}
In the following part and as depicted in Fig.~\ref{fig:Fig8}, we present the different orders of the Fraunhofer diffraction patterns corresponding to various OAM numbers relative to the atomic interaction length ($L/\xi$).
%_________________
\begin{figure}[b]
\includegraphics[width=1\textwidth]{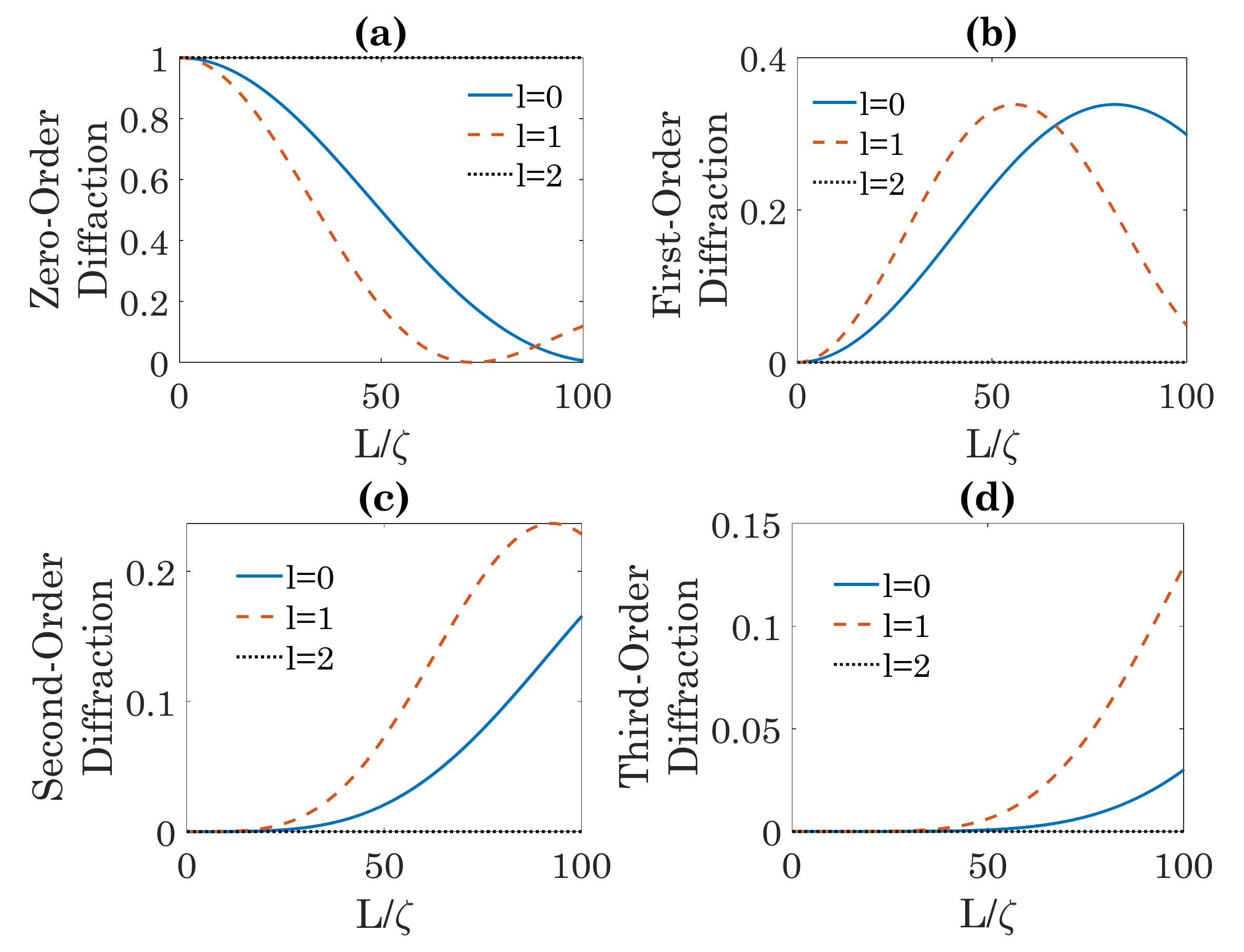}
\caption{\label{fig:Fig8} Different orders of Fraunhofer diffraction as a function of the atomic interaction length $L/\xi$ corresponds to (a)  the zero-order, (b) the first-order, (c) the second-order, and (d) the third-order for  $r/w=1$, $\varphi=\pi/4$ and $\Delta_p=\Delta_{c}=\Delta_{LG}=0$. It is fascinating to observe that the first to third-order contributions are prominent at larger values of $L/\xi$, while they do not have any significant impact at smaller $L/\xi$. The selected parameters are the same as in Fig. ~\ref{fig:Fig4}.}
\end{figure}
%______________________
In our analysis, we consider a composite vortex light with $r/w=1$ and $\varphi=\pi/4$. Additionally, all the incident lights are assumed to be in resonance with their respective transitions, i.e. $\Delta_p=\Delta_{c}=\Delta_{LG}=0$.
For the case of $l=0$ (solid line), all the intensity of the probe field accumulates at zero-order. However, as the interaction length $L$ increases, the zero-order intensity decreases and eventually reaches zero at $L=100\xi$. It is worth noting that the intensity of this order can be controlled by adjusting the value of the OAM number.
For $l=1$ (dashed line), the intensity of the zero-order also decreases, similar to the previous case. However, this time its intensity becomes zero at $L\simeq70\xi$.
In contrast, for $l=2$ (denoted by a dotted line), the entire intensity of the probe field remains at the zero-order, and its value does not change with varying the interaction length. This occurrence is because, for $l=2$, the intensity of the composite vortex light becomes zero (e.g. $\Omega_{LG}=0$), effectively converting the $N$-type atomic system into two independent two-level atomic systems. Under such conditions, the SW field has no effect on the Fraunhofer diffraction pattern of the probe field, resulting in intensities of the higher orders having initial zero values.
As the interaction length increases, the intensity of different orders grows, and this can be controlled by adjusting the value of the OAM number of the composite vortex field.
For the case of $l=0$ (solid line), with the decrease of the zero-order intensity, most of the intensity is transferred to the first order. When $L=100\xi$, the intensity of the zero-order has vanished, and all the energy is transferred to the first-, second-, and third-orders.
On the other hand, increasing the OAM to $l=1$ (dashed line) leads to most of the probe intensity gathering in the second and third orders.
\\
\indent
Gratings of various types have garnered significant interest in a wide range of applications, including optical switching \cite{Brown2005}, routing, and multibeam splitters \cite{Chen2017}, shaping biphoton spectra \cite{Wen2010}, and generating stationary light pulses \cite{Bajcsy2003}. 
Gratings with varying periods are also commonly used to measure topological charges  \cite{Kai2015, Zheng2017}.
These lattices enable dynamic control over the intensity and spatial distribution of diffraction patterns, making them particularly valuable in the context of multichannel optical communication and multiparticle capture applications.
\\
\indent
Considering these diverse applications, our proposed model holds significant promise for several purposes. 
It can be employed for the measurement of topological charges in optical vortex light, enabling a better understanding of complex light-matter interactions. 
Moreover, it has the potential to be a valuable tool in optical light shaping technologies, offering practical advantages in applications such as optical switching and the dynamic control of diffraction patterns.
%____________________________
%____________________________
%____________________________
\section{\label{sec:conclusions}Conclusions}
In summary, we have investigated the spatially dependent diffraction orders of Electromagnetically Induced Grating (EIG) in a four-level $N$-type atomic system. 
To achieve this, we employed an excitation scheme that simultaneously interacted with a standing wave beam along the $x$ direction and a composite vortex. 
Through the use of analytical methods for solving the integral equation of Fraunhofer diffraction, we established a direct link between the different diffraction orders and the corresponding Bessel functions of those orders. 
Our numerical analysis revealed that when the OAM of the composite beam is set to zero, distinctive ring patterns emerge for the various diffraction orders. 
These ring-shaped regions exhibit different intensities, resulting from constructive or destructive interferences. 
As the winding numbers of the composite beam increase, the ring patterns undergo fascinating transformations, evolving into petal-like structures with varying sizes depending on the diffraction order. 
The behavior of the different orders of Fraunhofer diffraction patterns upon varying the atomic interaction length for different OAM numbers shows the distribution of the probe field energy.
Taking advantage of the additional degree of freedom provided by the coupling composite optical vortex, we have proposed a straightforward scheme for controlling the performance of the EIG. 
The direct control offered by our method makes it experimentally feasible in common atom optics setups, and it holds great promise for constructing photonic devices and other elements for quantum technology applications.
The ability to tailor and control EIGs using composite optical vortices opens up exciting possibilities for advancements in quantum optics and related fields.
%____________________________________________________________
\begin{acknowledgments}
T. K. was supported by Grant No. LV-LT-TW/2023/10 “Coherent Optical Control of Atomic Systems” by the Ministry of Education and Science of the Republic of Latvia.
H. R. H. acknowledges support from Grant No. S-LLT-22-2 “Coherent Optical Control of Atomic Systems” by the Lithuanian Council of Research.
\end{acknowledgments}

% whether or not they are referenced in the text. This is an appropriate
\nocite{*}


\begin{thebibliography}{54}

\bibitem{Ling1998}
H. Y. Ling, Y. Q. Li, and M. Xiao,
Electromagnetically induced grating: Homogeneously broadened medium,
\textit{Phys. Rev. A}
\textbf{57(2)},
1338,
DOI:{https://doi.org/10.1103/PhysRevA.57.1338}
(1998).

\bibitem{Mitsunaga1999}
M. Mitsunaga and N. Imoto,
Observation of an electromagnetically induced grating in cold sodium atoms,
\textit{Phys. Rev. A}
\textbf{59(6)},
4773,
DOI:{https://doi.org/10.1103/PhysRevA.59.4773},
(1999).
  

\bibitem{Cardoso2002}
G. S. Cardoso and J. W. R. Tabosa,
Electromagnetically induced gratings in a degenerate open two-level system,
\textit{Phys. Rev. A}
\textbf{65},
033803,
DOI:https://doi.org/10.1103/PhysRevA.65.033803,
(2002).


\bibitem{Harris1997}
S. E. Harris,
Electromagnetically induced transparency,
\textit{Phys. Today}
\textbf{50(7)}, 
36,
DOI:https://doi.org/10.1063/1.881806, 
(1997).
   

\bibitem{Fleischhauer2005}
   M. Fleischhauer, A. Imamo\^{g}lu and J. P. Marangos,
   Electromagnetically induced transparency: Optics in coherent media,
   \textit{Rev. Mod. Phys.}
   \textbf{77(2)},
   633,  
   DOI:https://doi.org/10.1103/RevModPhys.77.633,
   (2005).

\bibitem{Harris1990}
 S. E. Harris, J. E. Field, and A. Imamo\^{g}lu,
 Nonlinear optical processes using electromagnetically induced transparency,
 \textit{Phys. Rev. Lett.}
   \textbf{64(10)},
  1107,  
   DOI:https://doi.org/10.1103/PhysRevLett.64.1107,
   (1990).
   
\bibitem{Lukin2000}
   M. D. Lukin and A. Imamo\^{g}lu,
 Nonlinear optics and
quantum entanglement of ultraslow single photons,
   \textit{Phys. Rev. Lett.}
   \textbf{84(7)},
  1419,  
   DOI:https://doi.org/10.1103/PhysRevLett.84.1419,
   (2000).
   
\bibitem{Brown2005}
   A. W. Brown and M. Xiao,
   All-optical switching and
routing based on an electromagnetically induced absorption grating,
   \textit{Opt. Lett.}
   \textbf{30(7)},
   699,
   DOI:https://doi.org/10.1364/OL.30.000699,
(2005).

\bibitem{Moretti2010}
   D. Moretti, D. Felinto, J. Tabosa, and A. Lezama,
    Dynamics of a stored zeeman coherence grating in an external
 magnetic field,
   \textit{J Phys. B: At. Mol. Opt. Phys.}
    \textbf{43},
   115502,
   DOI:10.1088/0953-4075/43/11/115502,
(2010).

\bibitem{Zhao2010}
   L. Zhao, W. Duan, and S. F. Yelin,
    All-optical beam control
with high speed using image-induced blazed gratings
in coherent media,
   \textit{Phys. Rev. A}
   \textbf{82},
   013809,
   DOI:https://doi.org/10.1103/PhysRevA.82.013809,
(2010).


\bibitem{Wen2017}
F. Wen, W. Wang, I. Ahmed, H. Wang, Y. Zhang,
Y. Zhang, A. R. Mahesar, and M. Xiao,
    Two-dimensional
talbot self-imaging via electromagnetically induced lattice,
   \textit{Sci. Rep.}
   \textbf{7},
 41790,
   DOI:https://doi.org/10.1038/srep41790,
(2017).


\bibitem{Zhai2001}
P. W. Zhai, X. M. Su, and J. Y. Gao,
Optical bistability
in electromagnetically induced grating,
   \textit{Phys. Lett. A}
   \textbf{289},
 27,
   DOI:https://doi.org/10.1016/S0375-9601(01)00576-X,
(2001).

\bibitem{Zhang2015}
Y. Q. Zhang, Z. K. Wu, M. R. Belić, H. B. Zheng, Z. G. Wang, M. Xiao, and Y. P. Zhang,
Photonic Floquet topological insulators in atomic ensembles
   \textit{Laser Photon. Rev.}
   \textbf{9},
331,
   DOI:10.1002/lpor.201400428,
(2015).

\bibitem{Zhang2013}
Y. Zhang, C. Yuan, Y. Zhang, H. Zheng, H. Chen, C. Li,
Z. Wang, and M. Xiao,
Surface solitons of four-wave
mixing in an electromagnetically induced lattice,
   \textit{Laser Phys. Letts}
   \textbf{10},
055406,
   DOI:10.1088/1612-2011/10/5/055406,
(2013).

\bibitem{Zhang2011}
Y. Zhang, Z. Wang, Z. Nie, C. Li, H. Chen, K. Lu, and
M. Xiao,
Four-wave mixing dipole soliton in laser-induced
atomic gratings,
   \textit{Phys. Rev. Lett.}
   \textbf{106},
093904,
   DOI:https://doi.org/10.1103/PhysRevLett.106.093904,
(2011).

\bibitem{Dutta2006}
  B. K. Dutta and P. K. Mahapatra,
    Electromagnetically
induced grating in a three-level $\xi$-type system driven by
a strong standing wave pump and weak probe fields,
   \textit{J. Phys. B}
   \textbf{39},
 1145,
   DOI:https://doi.org/10.1088/0953-4075/39/5/013,
(2006).

\bibitem{Naseri2014}
  T. Naseri and R. Sadighi-Bonabi,
    Electromagnetically induced
phase grating via population trapping conditions in
a microwave-driven four-level atomic system,
   \textit{J. Opt. Soc. Am. B}
   \textbf{31},
2879,
   DOI:https://doi.org/10.1364/JOSAB.31.002879,
(2014).

\bibitem{Wang2014}
  L. Wang, F. Zhou, P. Hu, Y. Niu, and S. Gong,
    Two dimensional
electromagnetically induced cross-grating in
a four-level tripod-type atomic system,
   \textit{J. Phys. B: At. Mol. Opt. Phys.}
   \textbf{47(22)},
 225501,
   DOI:https://doi.org/10.1088/0953-4075/47/22/225501,
(2014).

\bibitem{Sahrai2016}
 M. Sahrai, F. Bozorgzadeh, and H. Khoshsima,
    Phase
control of electromagnetically induced grating in a four-level
atomic system,
   \textit{Opt. Quant. Electron.}
   \textbf{48},
 438,
   DOI:https://doi.org/10.1007/s11082-016-0713-9,
(2016).


\bibitem{Hu2020}
 Y. Hu, G. Cheng, and A. Chen,
   Tunneling-induced phase grating in quantum
dot molecules,
   \textit{Opt. Express}
   \textbf{28(20)},
 29805,
   DOI:https://doi.org/10.1364/OE.404566,
(2020).


\bibitem{Feili2022}
 M. M. Feili, A. Mortezapour, and A. A. Naemi,
   Phase-controlled electromagnetically induced grating in a quantum dot molocule,
   \textit{Int. J Theor. Phys.}
   \textbf{61},
 27,
   DOI:https://doi.org/10.1007/s10773-022-04978-2,
(2022).

\bibitem{Liu2022}
 Y. Liu, Y. Xiang, and A. A. Mohammed,
   Asymmetric diffraction grating via optical vortex light in a tunneling quantum dot molecule,
   \textit{Laser Phys. Lett.}
   \textbf{19(9)},
 095205,
   DOI:10.1088/1612-202X/ac81bb,
(2022).


\bibitem{Asadpour2018}
 S. H. Asadpour, A. Panahpour, and M. Jafari,
   Phase-dependent electromagnetically induced grating in
a four-level quantum system near a plasmonic nanostructure,
   \textit{Eur. Phys. J. Plus}
   \textbf{133},
 411,
  DOI:hhttps://doi.org/10.1140/epjp/i2018-12221-9
(2018).


\bibitem{Asadpour2023}
 S. H. Asadpour, T. Kirova, H. R. Hamedi, V. Yannopapas, and E. Paspalakis,
   Azimuthal dependence of electromagnetically induced grating in a double V-type atomic system near a plasmonic nanostructure,
   \textit{Eur. Phys. J. Plus}
   \textbf{138},
 246,
  DOI:https://doi.org/10.1140/epjp/s13360-023-03871-z,
(2023).

\bibitem{Asadpour2021}
 S. H. Asadpour, T. Kirova, J. Qia, H. R. Hamedi, G. Juzeli\=unas, and E. Paspalakis,
   Azimuthal modulation
of electromagnetically induced
grating using structured light,
   \textit{Sci. Reports}
   \textbf{11},
 20721,
  DOI:https://doi.org/10.1038/s41598-021-00141-9,
(2021).

\bibitem{Allen1999}
 L. Allen, M. J. Padgett, and M. Babiker,
   IV The orbital angular momentum of light,
   \textit{Prog. Opt.}
   \textbf{39},
 291,
  DOI:https://doi.org/10.1016/S0079-6638(08)70391-3,
(1999).

\bibitem{Padgett2004}
 M. Padgett, J. Courtial, and L. Allen,
   Light's orbital angular momentum,
   \textit{Phys. Today}
   \textbf{57},
 35,
  DOI:https://doi.org/10.1063/1.1768672,
(2004).

\bibitem{Babiker2019}
 M. Babiker, D. L. Andrews, and V. E. Lembessis,
   Atoms in complex twisted light,
   \textit{J. Opt.}
   \textbf{21},
 013001,
  DOI: 10.1088/2040-8986/aaed14,
(2019).

\bibitem{Dutton2004}
 Z. Dutton and J. Ruostekoski,
   Transfer and storage of vortex states in light and matter waves,
   \textit{Phys. Rev. Lett.}
   \textbf{93},
193062,
  DOI:https://doi.org/10.1103/PhysRevLett.93.193602,
(2004).


\bibitem{Hamedi2019A}
 H. R. Hamedi, J. Ruseckas, E. Paspalakis, and
   G. Juzeli\=unas,
   Transfer of optical vortices in coherently prepared media,
   \textit{Phys. Rev. A}
   \textbf{99},
033812,
  DOI:https://doi.org/10.1103/PhysRevA.99.033812,
(2019).


\bibitem{Ruseckas2011}
  J. Ruseckas, A. Mekys, and G. Juzeli\=unas,
   Optical vortices of slow light using a tripod scheme,
   \textit{J. Opt.}
   \textbf{13},
064013,
  DOI:10.1088/2040-8978/13/6/064013,
(2011).

\bibitem{Wang2008}
  T. Wang, L. Zhao, L. Jiang, and S. F. Yelin,
  Diffusion-induced decoherence of stored optical vortices,
   \textit{Phys. Rev. A}
   \textbf{77},
043815,
  DOI:https://doi.org/10.1103/PhysRevA.77.043815,
(2008).

\bibitem{Pugatch2007}
   R. Pugatch, M. Shuker, O. Firstenberg, A. Ron, and N. Davidson,
  Topological stability of stored optical vortices,
   \textit{Phys. Rev. Lett.}
   \textbf{98},
203601,
  DOI:https://doi.org/10.1103/PhysRevLett.98.203601,
(2007).


\bibitem{Moretti2009}
   D. Moretti, D. Felinto, and J. W. R. Tabosa,
  Collapses and revivals of stored orbital angular momentum of light in a cold-atom ensemble,
   \textit{Phys. Rev. A}
   \textbf{79},
023825,
  DOI:https://doi.org/10.1103/PhysRevA.79.023825,
(2009).


\bibitem{Maleev2003}
 D. Maleev, and G. Swartzlander,
 Composite optical vortices,
   \textit{J. Opt. Soc. Am. B}
   \textbf{20(6)},
1169,
   DOI:https://doi.org/10.1364/JOSAB.20.001169,
(2003).


\bibitem{Baumann2009}
 S. M. Baumann, D. M. Kalb, L. H. MacMillan, and E. J.
Galvez,
Propagation dynamics of optical vortices due to gouy phase,
   \textit{Opt. Express}
   \textbf{17(12)},
9818,
   DOI:https://doi.org/10.1364/OE.17.009818,
(2009).


\bibitem{Hamedi2019B}
 H. R. Hamedi, E. Paspalakis, G. Zlabys, G. Juzeliūnas, and J. Ruseckas,
Complete energy conversion between light beams
carrying orbital angular momentum using coherent population trapping for a coherently driven double-lambda atom-light-coupling scheme,
   \textit{Phys. Rev. A}
   \textbf{100},
023811,
   DOI:https://doi.org/10.1103/PhysRevA.100.023811,
(2019).


\bibitem{Asadpour2022}
 S. H. Asadpour, H. R. Hamedi, T. Kirova, and E. Paspalakis,
   Two-dimensional electromagnetically induced phase grating via composite vortex light,
   \textit{Phys. Rev. A}
   \textbf{105},
 043709,
  DOI:https://doi.org/10.1103/PhysRevA.105.043709,
(2022).

\bibitem{Wahab2023}
 A. Wahab, M. Abbas, and B. C. Sanders,
   Effect of composite vortex beam on a two-dimensional gain
assisted atomic grating,
   \textit{New J. Phys.}
   \textbf{25},
 053003,
  DOI:https://doi.org/10.1088/1367-2630/accc6e,
(2023).

\bibitem{Arkhipkin2023}
V. G. Arkhipkin, D. A. Ikonnikov, and S. A. Myslivets,
Diffraction of a Laguerre-Gaussian beam in Raman interaction with a spatially periodic pump field,
   \textit{Phys. Rev. A}
   \textbf{107},
023519,
   DOI:https://doi.org/10.1103/PhysRevA.107.023519,
(2023).

\bibitem{Ikonnikov2023}
D. A. Ikonnikov, S. A. Myslivets, V. G. Arkhipkin, and A. M. Vyunishev,
Near-field evolution of optical vortices and their spatial ordering behind a fork-shaped grating,
   \textit{Photonics}
   \textbf{10(4)},
469,
   DOI:https://doi.org/10.3390/photonics10040469,
(2023).

\bibitem{Franke2007}
S. Franke-Arnold, J. Leach, M. J. Padgett, V. E. Lembessis,
D. Ellinas, A. J. Wright, J. M. Girkin, P. \"{O}hberg,
and A. S. Arnold,
Optical ferris wheel for ultracold
atoms,
   \textit{Opt. Express}
   \textbf{15(14)},
8619,
   DOI:https://doi.org/10.1364/OE.15.008619,
(2007).

\bibitem{He2009}
X. He, P. Xu, J. Wang, and M. Zhan,
Rotating single
atoms in a ring lattice generated by a spatial light modulator,
   \textit{Opt. Express}
   \textbf{17(23)},
21007,
   DOI:https://doi.org/10.1364/OE.17.021007,
(2009).

\bibitem{Lembessis2017}
V. E. Lembessis, A. Alqarni, S. Alshamari, A. Siddig,
and O. M. Aldossary,
Artificial gauge magnetic and electric
fields for free two-level atoms interacting with optical
ferris wheel light fields,
   \textit{J. Opt. Soc. Am. B}
   \textbf{34(6)},
1122,
   DOI:https://doi.org/10.1364/JOSAB.34.001122,
(2017).

\bibitem{Blanchard2000}
P. M. Blanchard, D. J. Fisher, S. C. Woods, and A. H. Greenaway,
Phase-diversity wave-front sensing with a distorted diffraction grating,
   \textit{Appl.Opt.}
   \textbf{39(35)},
6649,
   DOI:https://doi.org/10.1364/AO.39.006649,
(2000).

\bibitem{Desyatnikov2005}
A. S. Desyatnikov, A. A. Sukhorukov, and Y. S. Kivshar,
Azimuthons: Spatially modulated vortex solitons,
   \textit{Phys. Rev. Lett.}
   \textbf{95},
203904,
   DOI:https://doi.org/10.1103/PhysRevLett.95.203904,
(2005).

\bibitem{Bekshaev2006}
A. Bekshaev and M. Soskin,
Rotational transformations and transverse energy flow in paraxial light beams: linear
azimuthons,
   \textit{Opt. Lett.}
   \textbf{31(14)},
2199,
   DOI:https://doi.org/10.1364/OL.31.002199,
(2006).

\bibitem{Brown2005}
A. W. Brown and M. Xiao,
All-optical switching and routing based on an electromagnetically induced absorption grating,
   \textit{Opt. Lett.}
   \textbf{30(7)},
699,
   DOI:https://doi.org/10.1364/OL.30.000699,
(2005).

\bibitem{Chen2017}
Y.-Y. Chen, Z.-Z. Liu, and R.-G. Wan,
Beam splitter and router via an incoherent pump-assisted electromagnetically induced blazed grating,
   \textit{Appl. Opt.}
   \textbf{56(20)},
5736,
   DOI:https://doi.org/10.1364/AO.56.005736,
(2017).

\bibitem{Wen2010}
J. Wen, Y.-H. Zhai, S. Du, M. Xiao,
Beam splitter and router via an incoherent pump-assisted electromagnetically induced blazed grating,
   \textit{Phys. Rev. A}
   \textbf{82},
043814,
   DOI:https://doi.org/10.1103/PhysRevA.82.043814,
(2010).

\bibitem{Bajcsy2003}
M. Bajcsy, A. S. Zibrov, and M. D. Lukin,
Stationary pulses of light in an atomic medium,
   \textit{Nature}
   \textbf{426},
638,
   DOI:https://doi.org/10.1038/nature02176,
(2003).

\bibitem{Kai2015}
K. Dai, C. Gao, L. Zhong, Q. Na, and Q. Wang,
Measuring OAM states of light beams with gradually-changing-period gratings,
   \textit{Opt. Lett.}
   \textbf{40},
562,
   DOI:https://doi.org/10.1364/OL.40.000562,
(2015).

\bibitem{Zheng2017}
S. Zheng and J. Wang,
 Measuring Orbital Angular Momentum (OAM) states of vortex Beams with annular gratings,
   \textit{Sci. Rep.}
   \textbf{7},
40781,
   DOI:https://doi.org/10.1038/srep40781,
(2017).


\end{thebibliography}

\end{document}